\documentclass[aps,prl,twocolumn,superscriptaddress,showpacs,floatfix,amsmath,amssymb]{revtex4-1}
\usepackage{graphicx} 
\usepackage{bm}
\usepackage{color,mathtools,bbm} 
\usepackage{wasysym}	
\usepackage{hyperref}  
\usepackage{lineno}
\usepackage{hypernat}
\usepackage{float} 
\usepackage{wrapfig}   
\usepackage{mciteplus}
\def\etal{{\em{et al}}}    
    
\DeclareRobustCommand{\specialp}{\ensuremath{\mathcal{P}}}    
\def\K{\mathbf{K}}
\def\F{\mathbf{F}}
\def\k{\mathbf{k}}
\def\tildek{\mathbf{\tilde k}}

\begin{document}

\title{Two-particle excitations under coexisting electron interaction and disorder}
\author{C. E. Ekuma}
\altaffiliation{George F. Adams Research Associate}
\email{cekuma1@gmail.com}
\affiliation{U.S. Army Research Laboratory, Aberdeen Proving Ground, MD 21005-5069}

\date{\today}
\begin{abstract}  
\noindent  
We study the combined impact of random disorder and electron-electron, and electron-hole interactions on the absorption spectra of a three-dimensional Hubbard Hamiltonian. We determine the single-particle Green's function within the typical medium dynamical cluster approximation. We solve the Bethe-Salpeter equation (BSE) to obtain the dynamical conductivity. Our results show that increasing disorder strength at a given interaction strength leads to decreased absorption with the dynamical conductivity, systematically going to zero at all frequencies, a fingerprint of a correlation-mediated electron localization. Surprisingly, our data reveal that taking into account the effects of electron-hole interactions through the BSE significantly changes the oscillator strength with a concomitant reduction in the critical disorder strengths $W_c^U$. We attribute this behavior to enhanced quantum correction induced by electron-hole interactions. 
\end{abstract}    

    
\maketitle 
\textit{Introduction}.--
Recent experiments have shown that most correlated materials contain a significant amount of defects, which appear to be intrinsic~\cite{PhysRevB.85.024504,Kramer1993,Soukoulis10.1088,PhysRevB.93.041403,PhysRevLett.110.037001,efros1985electron,PhysRevB.71.125104}. These inhomogeneities could significantly affect device performance. Most of the experimental transport data on disordered materials have defied explanation by the conventional transport theory. For instance, the phase diagram of the binary mixture of the correlated ferromagnetic metal SrRuO$_3$ (T$_C \approx$160 K)~\cite{10.1021/ic50043a023} and the band insulator SrTiO$_3$ (band gap $\approx$ 3.2)~\cite{Ekuma3700433} is still under active research. One suggestion is that there is an Anderson insulator around $x\gtrapprox0.5$ and a disordered correlated insulator at $\sim$0.2~\cite{PhysRevB.71.125104}. Other potential candidates for which the coexistence of defects and electron-electron and electron-hole interactions could play a crucial role are the perovskite transition metal oxides, e.g., $A_{1-x}BA'_xO_3$. Understanding the defect morphology could greatly improve better characterization of their properties and that of materials in general.

There is a decade of history of theoretical research into electron localization. The majority of these computational/theoretical works focus on localization due to disorder or electron-electron interactions~\cite{Vollhardt_Wolfle,0022-3719-12-7-018,PhysRevB.50.1430,Mirlin-RMP,Kramer1993,50years,PhysRevB.30.1686,PhysRevB.31.6483,PhysRevB.32.7811,PhysRevB.36.8649,PhysRevB.23.1304}. These two limiting cases were pioneered by Anderson~\cite{Anderson1958,gang4} and Mott~\cite{Mott1967,Mott}, now known as Anderson and Mott localization, respectively. As explained above, defects and electron-electron interactions coexist in many physical systems and they can both be substantial. Also, in some cases, due to dynamical screening in the local environment of the system, the transport is no longer driven by electron or hole carriers but dominated by bound electron-hole pairs known as excitons. One consequence of this is the emergence of nontrivial many-body effects, e.g., spectral weight redistribution,  and multiferroicity~\cite{10.1021/jz3012436,doi:10.1021/nl0721113,5887374,Baldini2017} not observed in conventional systems. The incipient of electron localization in an otherwise ``strongly'' correlated system is generally difficult to model due to the competing energy scales that abound in this regime. Based on model-coupling theory, G\"{o}tze~\cite{0022-3719-12-7-018,10.1080/13642818108221896} developed a self-consistent localization formalism, which has been used by many authors, e.g., Prelov\v{s}ek~\cite{PhysRevB.23.1304} to calculate the conductivity of the noninteracting electron system. An approach based on the potential well analogy of the coherent potential approximation was formulated and used to calculate the conductivity of various disorder distributions~\cite{PhysRevB.30.1686, 
PhysRevB.31.6483,PhysRevB.32.7811,PhysRevB.36.8649,Soukoulis2003,Soukoulis10.1088}. The diagrammatic, self-consistent approach of Vollhardt and W\"{o}fle~\cite{Vollhardt_Wolfle} was used to calculate conductivity for the Anderson model~\cite{0022-3719-21-36-009,0022-3719-17-11-012,Kotov1983raey,10.1143JPSJ.55.3991}, and various other  models~\cite{PhysRevB.30.1686,PhysRevB.32.7811,PhysRevB.34.2253,PSSB2221380154}. Aguair \etal~\cite{Aguiar2013} used the inverse of the typical density of states as an approximation to the resistivity and showed that the resistivity curves as a function of temperature are reminiscent of the Mooij correlations originally observed in disordered transition metal alloys~\cite{PSSA2210170217}. Girvin and Jonson~\cite{PhysRevB.22.3583} introduced an approximate scheme for calculating conductivity that becomes accurate only close to the localization transition. This method was further used by Dobrosavljevi\'{c} \etal~\cite{Vlad2003} in their study using Bethe lattice. Zhang \etal~\cite{zhang2017} proposed a two-particle formalism and calculated the dc-conductivity within the typical medium theory for the noninteracting fermionic system. 

In this paper, we present and explore the absorption properties of a disordered Hubbard model at experimentally relevant Hubbard interactions using the typical medium dynamical cluster approximation~\cite{PhysRevLett.118.106404,PhysRevB.89.081107,PhysRevB.92.201114,PhysRevB.92.014209,0953-8984-26-27-274209,PhysRevB.92.205111,PhysRevB.94.224208}. Herein, we focus on the limit where the disorder and the kinetic term are far greater than the interaction strength (i.e., the interaction strength is far smaller than the noninteracting bandwidth). Also, we will explore the regime where disorder and interaction strength are both substantial (i.e., the interaction strength is large but still significantly smaller than the noninteracting bandwidth). The former is reminiscent of a correlated and strongly disordered semiconductor, e.g., Si:B~\cite{PhysRevLett.75.4266} and the latter could be compared to the perovskite compounds, e.g., (Ca,Sr)VO$_3$. We will, however, not explore the Mott physics, which is in the regime where the interaction strength is far greater than the effective bandwidth. This regime has been extensively studied in the literature (see, e.g., Refs~\cite{PhysRevLett.62.324,PhysRevB.51.10411,Hirschfeld-review,RevModPhys.66.261,RevModPhys.68.13,PhysRevLett.74.1178}). The main finding of this work is that electron-hole interactions significantly alters the critical behavior of a disordered, three-dimensional Hubbard Hamiltonian. Our calculations reveal that the critical disorder strengths are reduced by more than 10\% due to electron-hole interaction effects.

It is worthwhile to contrast the method presented herein with other approaches of calculating absorption spectra~\cite{Vollhardt_Wolfle,0022-3719-12-7-018,PhysRevB.50.1430,Mirlin-RMP,Kramer1993,50years}. The single-particle Green's functions used in our two-particle calculations are obtained self-consistently from a mean-field approach with an intrinsic order parameter for characterizing electron localization even in the proximity of a localization transition~\cite{PhysRevLett.118.106404,PhysRevB.89.081107,PhysRevB.92.201114}. Our approach also takes into account resonance effects, which systematically incorporate longer-range spatial fluctuations up to the system (cluster) size. This resonance effect is due to having more than one lattice site in the system as opposed to just one impurity site, e.g., as in the coherent potential approximation. The carriers now collide with each other as well as scatter off multiple lattice sites. One consequence of this inter-site correlation effect is coherent backscattering, which is a precursor to Anderson localization in a disordered system. We further take into account vertex corrections within the cluster. The vertex correction accounts for the polarization effects in the effective medium beyond the leading order of the perturbation theory (see, e.g., Refs.~\cite{gross1999relativistic,Gross1991,Negele,fetter1971quantum}. The typical medium, inter-site correlations, and the vertex corrections ensure proper characterization of the large fluctuations in the local Green's function that could lead to its \textit{typical value} being far removed from the average one~\cite{Miranda2012,PhysRevB.92.014209}. Unless otherwise stated, all the results presented herein are for the three-dimensional cubic lattice with a size of $3\times 3 \times 3$, corresponding to a cluster size $N_c=27$. We will focus on the paramagnetic phase, i.e, we do not allow for the formation of any local moments. All the reported results are obtained at zero temperature. We used a broadening parameter of 10$^{-4}$ and a computational accuracy (numerical uncertainty) of up to $\sim \pm 0.1$ in our calculations.  

\textit{Method}.--
We consider the Hubbard Hamiltonian of interacting electrons subjected to quenched random disorders 
\begin{equation}    
  \label{eqn:model}    
H=-\sum_{\langle i j \rangle \sigma}t_{ij}(c_{i \sigma}^{\dagger}c^{\phantom{\dagger}}_{j \sigma}+h.c.)+\sum_{i} U_i n_{i \uparrow} n_{i \downarrow} + \sum_{i \sigma} V_in_{i \sigma},
\end{equation}
where the first term describes the hopping of electrons on the lattice, the second term describes the energy cost of having two electrons with opposite spin sitting on the same lattice site, and the last term depicts the disorder potential. Herein, $c_{i}^\dagger$($c^{\phantom{\dagger}}_{i}$) is the creation (annihilation) operator of an electron on site $i$ with spin $\sigma$, $n_{i} = c_{i}^\dagger c^{\phantom{\dagger}}_{i}$ is the number operator, $t_{ij}=t$ is the hopping matrix element between nearest-neighbor sites, and $U_i=U$ is the electron-electron interactions strength parameterized by the Hubbard onsite energy. The disorder is represented by a spatially, uncorrelated, spin-independent random potential $V_i$ distributed according to a probability distribution function $P(V_i)=\frac{1}{2W}\Theta(W-|V_i|)$, where $\Theta(x)$ is the Heaviside step function and $W$ is the width of the box, which parametrizes the strength of the disorder. We set the energy units to $4t$. 

To calculate the two-particle Green's function, we need the single-particle counterpart. To obtain the single-particle Green's function $G(\vec k,E)$ in the presence of electron-electron interactions and random disorder, we solved the typical medium dynamical cluster approximation (TMDCA) self-consistency equations. The TMDCA maps the lattice problem (\ref{eqn:model}) onto a periodically repeated cluster of size $N_c$ primitive cells embedded in a typical medium. This typical medium is characterized by a self-consistently determined non-local, hybridization function $\Delta(\vec k,E)$ ~\cite{PhysRevLett.118.106404,PhysRevB.89.081107,PhysRevB.92.201114,PhysRevB.92.014209,0953-8984-26-27-274209,PhysRevB.92.205111,PhysRevB.94.224208,Jarrell01,PhysRevB.61.12739,RevModPhys.77.1027,PhysRevLett.69.168,EkumaDissertation2015}. The mapping is accomplished by dividing the first Brillouin zone of the original lattice into $N_c$ nonoverlapping equal cells. As one increases $N_c$, longer-range spatial fluctuations are systematically accounted for up to $\lesssim N_c^{1/d}$, where $d$ is the spatial dimension. The TMDCA self-consistency could be summarized as follows. We make an initial guess of hybridization function; $\Delta(\vec k,E)$ describes how the cluster sites couple to the typical medium. Using $\Delta(\vec k,E)$, we calculate the fully dressed cluster Green's function $G^c(E)=(\mathcal G^{-1}-V-\Sigma^\mathrm{Int})^{-1}$, where $\mathcal G$ is the cluster-excluded Green's function, $V$ is the disorder potential, and $\Sigma^\mathrm{Int}$ is the electron-electron interactions, which is included up to its second-order perturbation expansion. We note that the disorder is accounted for exactly within the cluster and $\Sigma^\mathrm{Int}$ is obtained self-consistently within the cluster solver using second-order perturbation theory. The cluster density of states $\rho^c=-\frac{1}{\pi} \mathfrak{Im}\,G^c$ is then calculated by averaging over a large number of configurations to obtain the momentum dependent, non-self-averaged typical density of states ~\cite{PhysRevB.89.081107,PhysRevB.92.201114,PhysRevB.92.201114}
\begin{equation}
    \rho_{t}^c(\K) = \langle \rho^c_i\rangle_\mathrm{geom} \left\langle \frac{\rho^c(\K)}{\frac{1}{N_c} \sum_{i}\rho_{i}^c}\right\rangle_\mathrm{arit},
    \label{e.2}
\end{equation}
where $\langle \rho^c_i\rangle_\mathrm{geom}=\exp\,\langle\ln\rho_i\rangle_\mathrm{arit}$ is the diagonal elements of $\rho^c$ and the second factor ensures that non-local fluctuations up to  $\lesssim N_c^{1/d}$ are captured within the typical environment. Using the Hilbert transformation, we obtain the cluster typical Green's function $G_t^c(\K)$ from $\rho_{t}^c(\K)$ and then calculate the coarse-grained Green's function 

\begin{equation}
    \bar{G} (\K) = \frac{N_c}{N} \sum_{\tildek}  \bigg [\displaystyle G^c_{t} (\K)^{-1} + \Delta (\K) - \epsilon(\k) + \bar{\epsilon}(\K) + \mu \bigg ]^{-1},
    \label{e.3}
\end{equation}
where the overbar depicts cluster coarse-graining and $\mu$ is the chemical potential. The TMDCA loop is closed by calculating a new hybridization function
\begin{equation}
    \Delta_\mathrm{n}(\K) = (1-\varsigma)\Delta_\mathrm{o}(\K)+ \xi \left[(G^c)^{-1}-\bar G^{-1}\right],
    \label{e.4}
\end{equation}
where $\Delta_\mathrm{n}$ ($\Delta_\mathrm{o}$) refers to the new (old) hybridization function and $\xi$ is a mixing parameter. Convergence is achieved when $G_t^c \approx \bar{G}$, which also coincides with $\Delta_\mathrm{n} \approx \Delta_\mathrm{o}$.

To determine the two-particle properties of the many-body Hamiltonian above, we solve the Bethe-Salpeter equation using the converged, single-particle Green's function obtained from the above TMDCA self-consistency equations as input. Herein, we focus on the particle-hole channel and calculate the dynamical, conductivity with and without electron-hole interactions. We obtain the full lattice, dynamical conductivity by solving the Bethe-Salpeter equations as outlined below.  
\begin{enumerate}
\item 
The TMDCA self-consistency equations are solved to obtain the single-particle Green's functions used in the two particle calculations. This requires both the single particle retarded $G^R(\vec k,E)$ and advanced $G^A(\vec k,E)$ Green's functions. However, since $A(\vec k,E)A(\vec k,E) = \frac{1}{2\pi i} \left [ \vartheta \right ] \times  \frac{1}{2\pi i} \left [ \vartheta\right ]$, where $\vartheta=G^A(\vec k,E)- G^R(\vec k,E)$ and $A(\vec k,E) = -\frac{1}{\pi} \mathfrak{Im}\,G(\vec k,E)$ is the spectral function, we require knowing only the retarded Green's function. In calculating the two-particle Green's function, we have used the averaged lattice and cluster Green's functions obtained within the typical medium. This is important as the underlying dynamics present in the system are encoded in these average quantities. Further and most importantly, these averaged quantities are the only ones that represent the physical Green's functions of the material. In the Matsubara frequency, the bare dynamic charge susceptibility $\chi_0(\vec q,i\omega)$ is 
\begin{equation}
\label{e.5m}
 \chi_0(\vec q, i\omega) = \frac{1}{\beta N} \sum_{\vec k, i E} \,G(\vec k+\vec q,iE+ i \hbar\omega)\,G(\vec k,iE)
\end{equation}
where $\beta$ is the inverse temperature~\cite{Negele,fetter1971quantum}. Generally, one needs to carryout analytic continuation of Eq.~\ref{e.5m} in order to calculate any observable. This process especially for disordered systems could miss important features in the spectra if not done carefully. However, since our cluster problem is solved in real space, we can avoid the analytic continuation by converting the Matsubara sums to real frequency integrals using spectra representation: $ G(\vec q, i\omega)=\int d \epsilon A[G](\epsilon)/(i\omega - \epsilon)$, where $A(\vec k,E) = -\frac{1}{\pi} \mathfrak{Im}\,G(\vec k,E)$ is the spectral function. Using the spectral representation, the Matsubara sum in Eq.~\ref{e.5m} could be converted to real frequency integrals as~\cite{Ekuma2018}   
\begin{widetext}
\begin{subequations}
\begin{align}
    \mathfrak{Im}\,\chi_0(\vec q,\omega) &= - \frac{2\pi}{N_c} \sum_{\vec k} \int_{-\infty}^{+\infty} [f(E)-f(E+\hbar\omega)]\,A(\vec k+\vec q,E+\hbar\omega)\,A(\vec k,E)\,\mathrm{d}E, \\
    \mathfrak{Re}\chi_0(\vec q,\omega) &= \frac{1}{\pi}\specialp \int_{-\infty}^{+\infty}\frac{\mathfrak{Im}\,\chi_0(\vec q,\omega)\mathrm{d}\omega'}{\omega'-\omega},
    \label{e.5}
\end{align}
\end{subequations}
\end{widetext}
where $\specialp$ denotes the principal value and $f(E)$ is the Fermi function. 

\item
The bare charge susceptibility for both the cluster ($c$) and the lattice ($l$) is then obtained as the renormalized one due to the screening within the typical medium as 
\begin{equation}
 \tilde \chi_0^{c/l}(\vec q,\omega)=\chi_0^{c/l}(\vec q,\omega)\left[\mathbbm{1}-U\chi_0^{c/l}(\vec q,\omega)\right]^{-1}
 \label{e.6}
\end{equation}
where $\mathbbm{1}$ is the identity matrix.

\item 
The lattice irreducible vertex is approximated with the cluster counterpart, i.e., $\Gamma^l \approx \Gamma^c \equiv\Gamma$~\cite{PhysRevB.61.12739,RevModPhys.77.1027,PhysRevLett.69.168,Jarrell01,EkumaDissertation2015}. The full lattice vertex function is then calculated using $\Gamma(\vec q,\omega)$ as
\begin{equation}
 \F(\vec q,\omega) = \Gamma(\vec q, \omega) [\mathbbm{1} -  \tilde \chi_0(\vec q,\omega)\Gamma(\vec q,\omega)]^{-1}
 \label{e.7}
\end{equation}

The full vertex function $\F$ includes all the possible scattering events between any two propagating particles. Diagrammatically, $\F$ consists of all the fully connected two-particle diagrams to infinite orders and is, as such, reducible. On the other hand, the irreducible vertex function $\Gamma$ is a subclass of the two-particle diagram in $\F$ that cannot be separated into two distinct parts by cutting two internal Green's function lines in any given channel~\cite{PhysRevB.61.12739,RevModPhys.77.1027,Jarrell01}.

\item 
With the full lattice vertex function and the renormalized dynamical charge susceptibility calculated, the full, dynamical lattice susceptibility is obtained
\begin{align}
    \langle \vec k|\chi|\vec k'\rangle &= \langle \vec k|\tilde \chi_0|\vec k\rangle+\sum_{\vec k''}\langle \vec k|\tilde \chi_0|\vec k\rangle\langle \vec k|\F|\vec k''\rangle\langle \vec k''|\chi|\vec k'\rangle.
    \label{e.8}
\end{align}

\item 
The real part of the dynamical conductivity that takes into account electron-hole interactions (exciton) effects $\sigma_{eh}(\omega)$ is then obtained from Eq.~\ref{e.8} as 
\begin{equation}
 \label{e.8a}
 \sigma_{eh}(\omega) = \lim\limits_{\vec q \to 0} \frac{1}{\omega} \mathfrak{Im} \chi (\vec q,\omega)
\end{equation}

\end{enumerate}

\begin{figure}[b!]
    \includegraphics[trim = 0mm 0mm 0mm 0mm,scale=0.5,keepaspectratio,clip=true]{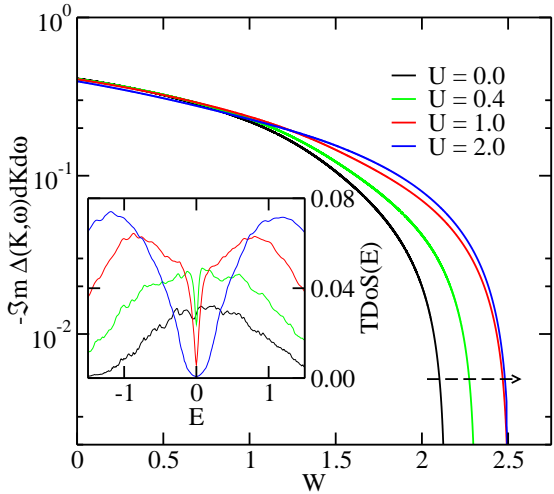}
    \caption{The semi-log plot of the integrated, imaginary part of the hybridization function for a $3 \times 3 \times 3$ cubic lattice sites at various interaction strengths, $U=0.0$, 0.4, 1.0, and 2.0 in the units of 4$t$. The arrow indicates the systematic increase of the critical disorder strengths $W_c^U$ due to interaction induced delocalization (disorder screening). The obtained $W_c^U$ are 2.13, 2.21, 2.51, and 2.49 for the $3 \times 3 \times 3$ cubic lattice sites. The inset is the typical density of states for $U=0.0$, 0.4, 1.0, and 2.0 at $W=2.0$, which is close to the $W_c^{U=0.0}$. An unconventional soft-pseudogap develops at the Fermi level for small interaction $U\ll W$, which systematically evolves into a conventional hard-gap at large $U$. The former gap is linear in $E$ while the latter is $E^2$-dependent. Observe that this gap is absent for $U=0$.}
    \label{f.1}
\end{figure}

\textit{Results}.--
We start the discussion of our results by presenting in Fig~\ref{f.1} the single-particle quantity as manifested in the imaginary part of the integrated hybridization function $\mathfrak{Im} \int \Delta(\K,\omega) d\K d\omega$ for various disorder and interaction strengths, respectively. The hybridization function is a natural order parameter for characterizing disordered systems as it measures the probability of how the electrons move between the cluster and the host (escape rate)~\cite{PhysRevB.92.014209}. In the dilute limit, i.e., small disorder strength up to $W\approx 0.5$, the hybridization function is practically the same for all the interaction strengths studied. However, as the strength of the disorder increases and in the limit where the interaction strength is far smaller than the noninteracting bandwidth of 3 (in unit of $4t$), the spectra starts to deviate from each other with the critical disorder strength $W_c^U$ systematically moving to higher values (as indicated by the arrow) for increasing interaction strength. Observe also that as both $W$ and $U$ becomes substantial and comparable to each other, the delocalization of the states rather increases. This is different from the monotonic decrease in the magnitude of the spectra for increasing disorder strength observed in the noninteracting systems~\cite{Bulka85,Slevin99,PhysRevLett.105.046403,Schubert2005,PhysRevB.76.045105,
PhysRevB.63.045108,PhysRevB.84.134209,PhysRevLett.47.1546,PhysRevB.92.014209,PhysRevB.89.081107}. The renormalization of the spectra and the increase in $W_c^U$ could be attributed to delocalization induced by $U$, which injects mobile carriers into the system. This is in agreement with the conclusions reached using the typical density of states as an order parameter~\cite{PhysRevB.92.201114} and has been interpreted by various authors to be due to disorder screening~\cite{Ma1982,PhysRevLett.91.066603,Aguiar2013,PhysRevB.92.201114}. Our calculations for the various interaction strengths of 0.0, 0.4, 1.0, and 2.0 also led to $W_c^U$ of 2.13, 2.21, 2.51, and 2.49. The critical disorder strength of 2.13 for the noninteracting limit is in good agreement with that obtained using the typical density of states within the typical medium dynamical cluster approximation~\cite{PhysRevB.92.201114,PhysRevB.89.081107,PhysRevB.92.201114} and with the numerically exact value $W_c\approx2.10$~\cite{Bulka85,Slevin99,PhysRevLett.105.046403,Schubert2005,PhysRevB.76.045105,
PhysRevB.63.045108,PhysRevB.84.134209,PhysRevLett.47.1546}. 

\begin{figure}
    \includegraphics[trim = 0mm 0mm 0mm 0mm,scale=0.5,keepaspectratio,clip=true]{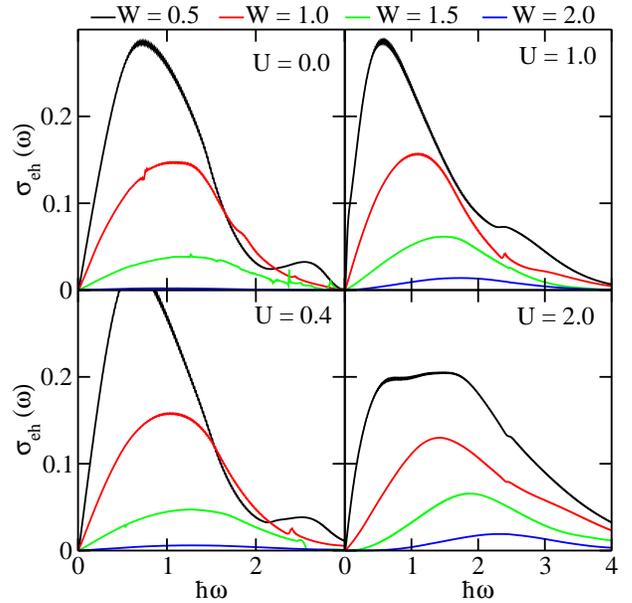}
    \caption{The dynamical conductivity obtained with the effects of electron-hole interactions included for a disordered Hubbard model as a function of the excitation energy $\hbar\omega$ obtained using Eq.~\ref{e.8}. Electron-hole interactions effects are included using the Bethe-Salpeter equations. The disorder strengths are 0.5, 1.0, 1.5, and 2.0 for the Hubbard interaction strengths $U=0.0$, 0.4, 1.0, and 1.0, respectively in the units of 4$t$.}
    \label{f.2}
\end{figure}  

The inset in Fig~\ref{f.1} shows the typical density of states obtained at $W=2.0$ for various interaction strengths. This disorder strength is close to the $W_c^{U=0.0}$ and could be said to depict a strongly disordered system. Observe that at $U=0.0$, there is no gap in the spectra. However, for finite $U$, a gap (which is independent of filling) opens at the Fermi level. For small $U$, this gap is an unconventional soft-pseudogap, which is almost linear in energy $E$. We have recently demonstrated that this soft-pseudogap emerges due to the reduction in phase space for scattering by $U$ and it is linear instead of the normal $E^2$-dependence due to the loss of momentum conservation~\cite{PhysRevB.92.201114}. Hence, a strongly disordered, correlated system ($W\gg U$) could be said to exhibit a non-Fermi liquid behavior since a well-defined quasiparticle could be said to no longer exist~\cite{GOLUBEV1998164,*refId0}. The deviation from the usual $E^2$ behavior in the vicinity of electron localization has been experimentally observed in some perovskite materials, e.g., $A_{1-x}BA'_xO_3$~\cite{PhysRevLett.80.4004,PhysRevB.73.235109,PhysRevB.76.165128}. For example, the photoemission spectra of SrRu$_{1-x}Ti_xO_3$ exhibit a soft pseudogap gap at $x=0.5$ and a hard gap at higher values of $x$~\cite{PhysRevB.76.165128}. Observe further from the inset that the soft-pseudogap systematically evolves into a hard-gap at large $U$ with the usual $E^2$-dependence behavior restored and inelastic scattering now vanishes as $E \rightarrow 0$, reminiscent of a Fermi liquid. This latter observation suggests that a strongly correlated and disordered system could be described using the Fermi liquid physics but the contrary may not be the case especially in the regime of strong disorder and weak interaction strength as observed herein.      
  
Next, we consider the two-particle quantities for a disordered Hubbard model. We show in Fig.~\ref{f.2} the calculated dynamical conductivity $\sigma_{eh}(\omega)$, which accounts for the effects of electron-hole interactions obtained using Eq.~\ref{e.8a} for the same parameters as in Fig.~\ref{f.1}. This spectrum also included vertex corrections. The vertex correction effects renormalized the spectra, which is more significant at low-energy $\omega < 1.0$. While the vertex corrections have subtle effects, i.e., it increases the magnitude of the low-energy of the absorption spectra (not shown), our calculations show that non-local corrections are more important for the proper description of the absorption spectra of correlated, disordered systems. We note that the former effect could become significant, e.g., for the description of transport phenomena in Kondo systems~\cite{PhysRevB.83.073301}. 

\begin{figure}[b!]
    \includegraphics[trim = 0mm 0mm 0mm 0mm,scale=0.35,keepaspectratio,clip=true]{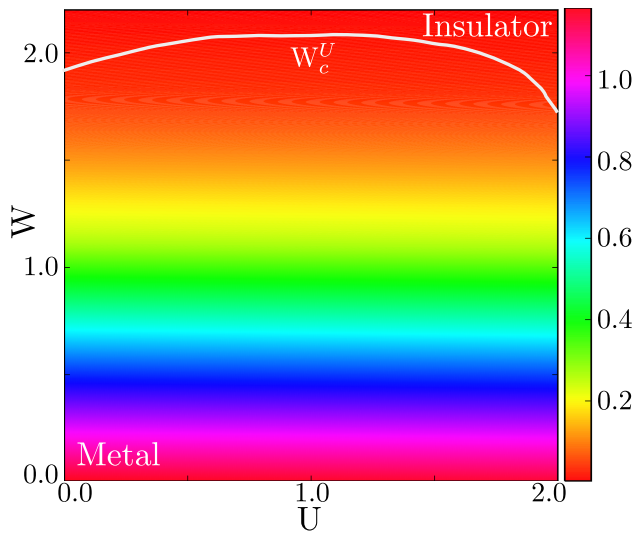}
    \caption{The contour plot of the disorder-interaction phase diagram of the dynamical conductivity obtained with the effects of electron-hole interactions included in units of $4t$. Data are obtained from Fig. ~\ref{f.2} by interpolating the maximum for each data set to a finer grid. The solid-white line is intended to give a rough estimate of the location of the critical disorder strengths in the parameter space. 
    The trend of W$_c^U$ is in agreement with previous studies~\cite{PhysRevLett.94.056404,PhysRevLett.102.156402}. }
    \label{f.3}
\end{figure}

Our data show different behaviors at different energies. At high-energy $\omega > 1.0$, we observed Lifshitz tails and the suppression of the spectra with significant broadening and a reduction in the oscillator strength. The latter measures the absorption probability. In the low-energy regime $\omega < 1.0$, observe that the Drude-like behavior normally observed at zero or small disorder strength (as can be seen in Fig.~\ref{f.3}) is absent. This can be understood by the transport now being dominated by the electron-hole pairs. The maximum of the spectra occurs at $\sim$ 1.0, and it is systematically blue-shifted as the strength of the disorder and interaction is increased. Our data further reveal that the initial delocalization effects are significantly higher at small $W$ and $U$. For example, the highest magnitude of the spectra occurs for the parameters $W=0.5$ and $U=0.4$. However, in the intermediate and strong disorder limit $1.5\leqq W=2.0$, the delocalization effects systematically increases as $U$ is increased. We explain this observation as follows: when the disorder strength is small and the interaction is finite but also smaller, more free electron-hole pairs are generated leading to the observed increase in conductivity. Still, even in the weak disorder limit, if the interaction strength is significantly larger than $W$, ``strongly correlated'' physics could dominate. The system adopts a Mott-like behavior preferring to open a gap at the Fermi level due to less generation of free electron-hole pairs (see the inset of Fig.~\ref{f.1} where increasing $U$ induces the opening of a gap at the Fermi level). On the other hand, when the interaction strength is large and the disorder strength is close to the noninteracting critical disorder limit, the system could become a correlated dirty metal leading to the observed delocalization in this regime.

Generally, the single- and two-particle behaviors are qualitatively similar since they both systematically go to zero as the strength of the disorder is increased. But quantitatively, significant differences exist in their critical behavior. For instance, the two-particle calculations led to critical disorder strengths that are far smaller than their single-particle counterparts. 

We can further gain some insights on how the critical quantities, e.g., the critical disorder strengths change in the two-particle picture by exploring the dc-conductivity, which can be obtained from the dynamical conductivity by taking the zero limit of the excitation energy as $\sigma_{eh}(\omega\rightarrow 0)$. In our analysis, we instead adopt the maximum value $\sigma_{eh}(\omega\rightarrow \omega_{max})$ to avoid any ambiguity due to the nature of the excitation spectra, e.g., Lifshitz tails~\cite{10.1080/00018736400101061}. The extracted $\sigma_{eh}(\omega\rightarrow \omega_{max})$ values were further interpolated to a finer grid. The associated contour plot is shown in Fig.~\ref{f.3}. The essence of this plot is to show more clearly, the overall evolution of the dynamical conductivity in the disorder-interaction parameter space. From Fig.~\ref{f.3}, up to $W\approx0.25$ for all $U$-values, we observed a `pure' metallic-like behavior. Then, we see a weakly interacting metallic character up to $W\approx0.75$, followed by some intermediate states, and then a correlated `dirty' metal before the system goes into the strongly correlated Anderson insulator regime. The  $W_c^U$ obtained from our data are: 1.93, 2.04, 2.11, and 1.98 for $U=0.0$, 0.4, 1.0, and 2.0, respectively. This shows a reduction of more than 10\%, e.g., $W_c^{U=0.0}$ is reduced by $\approx$ 0.2 when compared to the single-particle equivalent.

\begin{figure}[t!]
    \includegraphics[trim = 0mm 0mm 0mm 0mm,scale=0.5,keepaspectratio,clip=true]{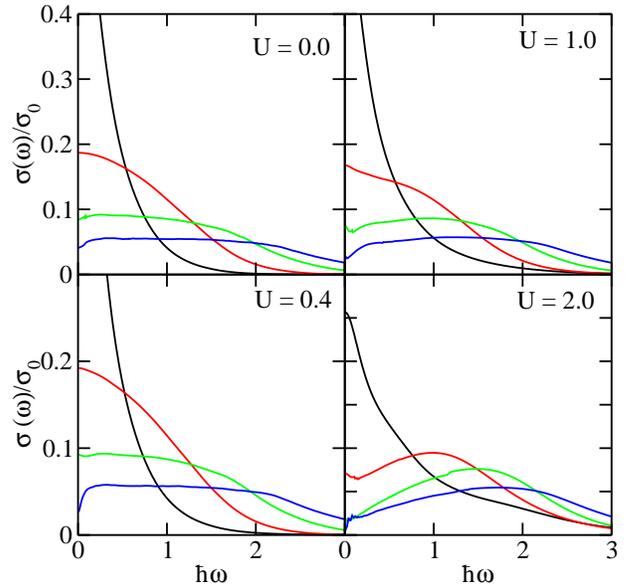}
    \caption{The dynamical conductivity obtained without the effects of electron-hole interactions ((normalized to its zero frequency value $\sigma_0$) at various interactions and disorder strengths for the same system size as in Fig.~\ref{f.1} obtained using Eq.~\ref{e.9} in units of $4t$. Observe the systematic evolution from Drude-like to non-Drude-like behavior and also the decrease in the oscillator strength in the low-energy regime as the strength of the disorder increases.}
    \label{f.4}
\end{figure}

To explore the origin of this discrepancy, we further calculated the dynamical conductivity (without the effects of electron-hole interactions) using the Kubo-Green'swood formula~\cite{10.1143/JPSJ.12.570,10.1142/S0217979211100977}
\begin{widetext}
\begin{equation}
\label{e.9}
 \sigma(\omega) = \frac{\sigma_0}{2 \pi^2} \mathfrak{Re} \int_{- \infty}^\infty \mathrm{d}E\, \frac{[f(E)-f(E+\omega)]}{\omega}\,  \left[\frac{G^*(E)-G(E+\omega)}{\gamma(E+\omega)-\gamma^*(E)} - \frac{G(E)-G(E+\omega)}{\gamma(E+\omega)-\gamma(E)} \right],  
\end{equation}
\end{widetext}
where $\gamma(\omega) = \omega -\mu-\Sigma(\omega)$ and $\sigma_0$ is the zero frequency value.
We show in Fig.~\ref{f.4} the plot of the dynamical conductivity obtained using Eq.~\ref{e.9} in unit of $\sigma_0$ for the same parameters as in Fig.~\ref{f.2}. Our results show a Drude-like behavior in the low-energy regime when the disorder strength is still small. However, for a given interaction, as the strength of the disorder increases, the conductivity is suppressed especially in the low-energy regime, which becomes non-Drude-like. At high-energy, the delocalization by interaction and the suppression of the spectra as $W$ increases are seen in both the single-particle hybridization function and the two-particle spectra data. We interpret this behavior as being due to quantum corrections to the Drude conductivity by both weak localization effects and the disorder-modified electron-electron interactions~\cite{Altshuler1985}.

As can be inferred from  Fig.~\ref{f.4}, the $\sigma(\omega \rightarrow 0)$ or $\sigma(\omega \rightarrow \omega_{max})$ is still significant at the disorder and interaction strengths where their two-particle counterpart that included the electron-hole interaction effects is already localized. For example, at $U=0.0$ and $W=2.0$, the calculated dynamical conductivity without the electron-hole interactions is still finite while the counterpart obtained from the Bethe-Salpeter equation is already practical zero. The overall trend of the critical parameters, e.g., $W_c^U$-values obtained in the absence of electron-hole interactions is in agreement with the ones calculated from the single-particle quantity. Since the critical behavior of the dynamical conductivity calculated without electron-hole interactions [Eq.~\ref{e.9}] is in basic agreement with the single-particle behavior of the critical quantities, we attribute the reduction in the critical disorder strengths in the presence of electron-hole interactions to enhanced multiscattering processes induced by the disorder, which breaks up the extended states within the system, leading to less generation of free electron-hole pairs. 

While we cannot directly verify the outcome with our data, the exciton states induce changes in the oscillator strengths, i.e., the relative heights/positions of the absorption spectra thereby lowering $W_c^U$. This is similar to what is observed in some materials in which an electron-hole pair has a binding energy that causes the quasiparticle gap to be higher than the fundamental gap obtained from conventional methods or measured via photoemission spectroscopy. Several experiments have shown that exciton effects drastically change the spectra of materials. The data of Varley and Schleife~\cite{0268-1242-30-2-024010} for some transparent conducting oxides showed that the absorption spectra are strongly modified by the inclusion of electron-hole interactions especially the lower photon-energy behavior which, was red-shifted. The redistribution of the spectral weight at low photon-energy due to excitonic effects was also reported for several oxides~\cite{PhysRevB.83.035116,PhysRevB.80.035112,schleife_bechstedt_2012}. The electron-hole interactions have also been demonstrated to be important in describing the properties of nanostructure materials, e.g., monolayer MoS$_2$ in which it is vital for the proper interpretation of the low-energy absorption spectra~\cite{PhysRevLett.111.216805} especially the position of the principal exciton peaks. The impact of electron-hole interactions could even be greater in disordered and/or interacting physical systems where the disorder degrees of freedom could couple nontrivially to the electron-electron interactions and/or the electron-hole interactions. Hence, the approach and the results presented herein could be of great importance in the understanding and interpretation of transport data of disordered and/or interacting systems where conventional approaches may not be adequate.

\textit{Summary}.--
We have presented and explored the role of electron-hole interactions in the disordered Hubbard model for a random disorder potential distributed according to a box probability distribution function in three dimensions using the typical medium approach. Our calculations reveal a significant reduction in the critical disorder strengths when compared to the single-particle values. We attribute this reduction in $W_c^U$ to enhanced coherent backscattering processes (cooperon correction) due to the inclusion of electron-hole interactions. 

\textit{Acknowledgment}.--
Research was sponsored by the Army Research Laboratory and was accomplished under the Cooperative Agreement Number W911NF-11-2-0030 as an ARL Research [George F. Adams] Fellow. This work was supported in part by a grant of computer time from the DOD High Performance Computing Modernization Program at the Army Engineer Research and Development Center, Vicksburg, MS.

\end{document}